\RequirePackage{fix-cm}
\documentclass[journal]{IEEEtran}       
\usepackage{graphicx}
\usepackage{amsmath}
\usepackage{fixltx2e}
\usepackage{float}
\usepackage[euler]{textgreek}
\usepackage{hyperref}
\usepackage{algorithmic}
\usepackage[ruled,vlined]{algorithm2e}
\begin{document}

\title{Continuous Lyapunov Controller and Chaotic Non-linear System Optimization using Deep Machine Learning}


\author{Amr~Mahmoud,~\IEEEmembership{}
        Youmna~Ismaeil~\IEEEmembership{}
        and~Mohamed~Zohdy~\IEEEmembership{}
\thanks{A. Mahmoud is with the Department of Electrical and computer engineering, Oakland University
Rochester, MI, 48307 USA e-mail: (amahmoud@oakland.edu).}
\thanks{Y. Ismaeil is with the Department of computer science, Saarland University
Saarbrücken,Germany e-mail: (s8yoisma@stud.uni-saarland.de).}
\thanks{M. Zohdy is with the Department of Electrical and computer engineering, Oakland University
Rochester, MI, 48307 USA e-mail: (mzohdy@oakland.edu).}}




\IEEEtitleabstractindextext{
\begin{abstract}
The introduction of unexpected system failures and nonlinear disturbances does not allow controllers to guarantee continuous system stability. In this research we present a novel approach for detecting early failure indicators in a non-linear highly chaotic system and accordingly predict the best parameter calibrations to offset such instability. The approach proposed utilizes deep machine learning regression model to continuously monitor and re-tune the system and controller signals. The Re-calibration of the parameters is triggered according to conditions designed to maintain stability without compromise to the system output stability, intended outcome or computation cost. The deep neural model works in parallel to the system to predict new parameters that would counteract expected system in-stability. The proposed approach is applied on a non-linear complex combination of Duffing Van der pol oscillators. The approach is also tested under different scenarios such as the system and controller parameters  being initialized incorrectly, the system actuator failure scenarios or nonlinear disturbances introduced while running to measure effectiveness and reaction time.
\end{abstract}
}
\maketitle
\IEEEdisplaynontitleabstractindextext

\section{Introduction}
\label{intro}
Lyapunov control has been proven successful in controlling highly chaotic non-linear oscillators \cite{Ref1}\cite{Ref2}\cite{Ref3} . One of the fundamentals that contribute to the success or failure of any type of control strategy is the controller and system parameters. Therefore, researchers have explored different methods to find the precise parameters that would lead to achieving the best system results \cite{Ref4}\cite{Ref5}. \\
One of the methods that was utilized to achieve the previously mentioned goals is Genetic algorithm (GA). GAs have been successful in cases where all the system dynamics are clearly defined and known to some extent or with systems where limited system disturbances are introduced and minor parameter tuning is required \cite{Ref6}\cite{Ref7}. In some cases, several system assumptions are needed in order to allow the GA to run successfully.\\ 

Due to some of the limitations found in using GAs such as inability to quickly converge to the final solution or adapt to unknown system dynamics or unknown disturbances. Researchers though after different approaches that wouldn't reduce the system agility and at the same time would be able to handle unknown system characteristics. A hybrid approach of Fuzzy Control and GAs was researched \cite{Ref8} but system linearization is a requirement in order to use the previously mentioned method .\\ 

Another approach that is recently being researched is the use of Machine Learning to enhance the controller performance. For example through the use of Episodic learning \cite{Ref9}\cite{Ref10}. Most recently, there is the introduction neural lyapunov control which proposes the use of deep learning to find the control and Lyapunov functions. The approach mentioned in \cite{Ref9}\cite{Ref10} is suitable for find the best system parameters that would initially lead the system to stability and reduced the system error. The problem with approach in \cite{Ref9}\cite{Ref10} is that it assumes that the system is deterministic, time invariant, and affine in the control input. while in real life situation external perturbations might occur resulting in system failure at any moment while the system is running \cite{Ref11}\cite{Ref12}. The approach proposed in \cite{Ref25} \cite{Ref26} and \cite{Ref27}attempts to predict the control and Lyapunov functions that would lead to system stability but under specific conditions where the system dynamics are deterministic in nature. The approach proposed in this research is novel to the best of our knowledge in that it discards the assumption of an ideal environment or fully known system dynamics and seeks continuous enhancement of the controller outcome through continuous monitoring of the system error, reference signal, system dynamics and control signal and accordingly adjust the system and controller parameters to improve the controller performance without the need to disrupt the system output.

 The focus of the research is to allow the Deep Learning Algorithm to learn the system from a continuously improving dataset and according to the slope of the output error the algorithm relearns the system and collects the needed information.\\   

The proposed method is applied to a non-linear chaotic combined system of Duffing and Van der pol oscillators\cite{Ref24}. The aforementioned system was chosen to test the Deep learning algorithm response to unpredicted system disturbances and unknown system dynamics\cite{Ref21}\cite{Ref22}\cite{Ref23}. An algorithm was developed to aid and trigger the Deep Neural Network when needed to adapt to new system dynamics and according to preset conditions. The algorithm records and feeds an updated data set to the DNN in order to relearn the system dynamics if certain conditions are detected to be true. Once the process of retraining is complete, the algorithm generates a random array of parameters and the array of parameters is fed to the DNN to predict the output error. If the DNN predicts that the error slope will be reduced from the current value improving the performance of the system, then the suggested controller parameters are allowed to be set as the new controller parameters. Otherwise the algorithm would create a new array set of randomly generated parameters and re-feed them to the DNN to provide its predictions in a continuous loop. Once the controller parameters are updated the algorithm monitors the actual error compared to the predicted error to determine the network viability and accordingly if the error difference is greater than a set value the algorithm would trigger a retraining request of the DNN. \\
\section{LYAPUNOV CONTROL ON DUFFING - VANDERPOL OSCILLATOR MODEL}
\label{Section II}
This section provides an overview of the Duffing-Van der pol system dynamics. 
Duffing-Van der pol mathematical model is  
\begin{equation}
\label{eq:1}
\ddot{x}+ \delta\dot{x}+ \epsilon_1( x^2\ddot{x}+x\dot{x}^2)+\epsilon_2x^3 = \mu(t) + \gamma cos(\omega t)	
\end{equation} 
It combines both non-linear stiffness and non-linear damping [1].
Setting
\begin{equation}
\label{eq:2}
\dot{x} = x_2  
\end{equation} 
and
\begin{equation}
\label{eq:3}
	 x = x_1
\end{equation}
therefore
\begin{equation}
\label{eq:4}
	 \dot{x_1} = x_2
\end{equation} 
we get the state space model
 \begin{equation}
\label{eq:5}
	 \dot{x_2} = \frac {1} {1+\epsilon_1 X^2} (\delta \dot{x} - \epsilon_1 x \dot{x^2} -  \epsilon_2 x^3 +\mu(t) +\gamma cos(\omega t)) 
\end{equation}   

\begin{equation}
\label{eq:6}
\begin{bmatrix}
\dot{x_1}  \\ 
\dot{x_2}  
\end{bmatrix}
 =
  \begin{bmatrix}
   x_2 \\
  \frac{1}{1+ \epsilon_1 x^2} (\delta x_2 - \epsilon_1 x x_2^2 - \epsilon_2x^3)  
   \end{bmatrix}
\end{equation}
\begin{equation}
\label{eq:7} 
y  = \begin{bmatrix}  x_1 \\  x_2     \end{bmatrix}
\end{equation}  
\section{LYAPUNOV CONTROLLER DESIGN}
\label{Section III}
In this section we give an overview of the controller design. The controller is designed with the purpose of achieving Duffing- van der pol oscillator system stability at an increased forcing frequency. Desired state q\textsubscript{d} and actual state q.
\begin{equation}
\label{eq:8}
e=q_d - q
\end{equation} 
A control LYAPUNOV candidate is then chosen to be
\begin{equation}
\label{eq:9}
V = \frac{1}{2} (\gamma_1 \dot{e} + \gamma_2 e)
\end{equation} 
As can be seen from \cite{Ref11}, the transient behaviour of the error dynamics can be influenced by a suitable choice of the CLF and gamma parameters, respectively \cite{Ref10}. Therefore the CLF has been chosen to be positive definite for all \begin{equation} x,\dot{x} \end{equation} taking the time derivative of V
\begin{equation}
\label{eq:11}
\dot{V} = (\gamma_1 \dot{e} + \gamma_2 e)(\gamma_1 \ddot{e} + \gamma_2 \dot{e})
\end{equation} 
\begin{equation}
\label{eq:12}
\dot{V} = -kV
\end{equation} 
which is globally exponentially stable as V is globally positive definite. Hence, by substitution in Eqn.\ref{eq:11} we find the control law
\begin{equation}
\label{eq:13}
u = \frac{1+\epsilon_1 x^2}{\gamma_2} (\frac{k}{2} \gamma_1 e+ \frac {k}{2} \gamma_2  \dot{e} + \gamma_1  \dot{e} + \gamma_2 \ddot{x_des}) - pCos(\omega t)+\delta \dot{x} + x + \epsilon_1  \dot{x} x^2 + \epsilon_2 x^3
\end{equation} 
\section{SIMULATION RESULTS FOR UNOPTIMIZED SYSTEM}
\label{Section IV}
\begin{figure}[!t]
\centering
\includegraphics[width=3in]{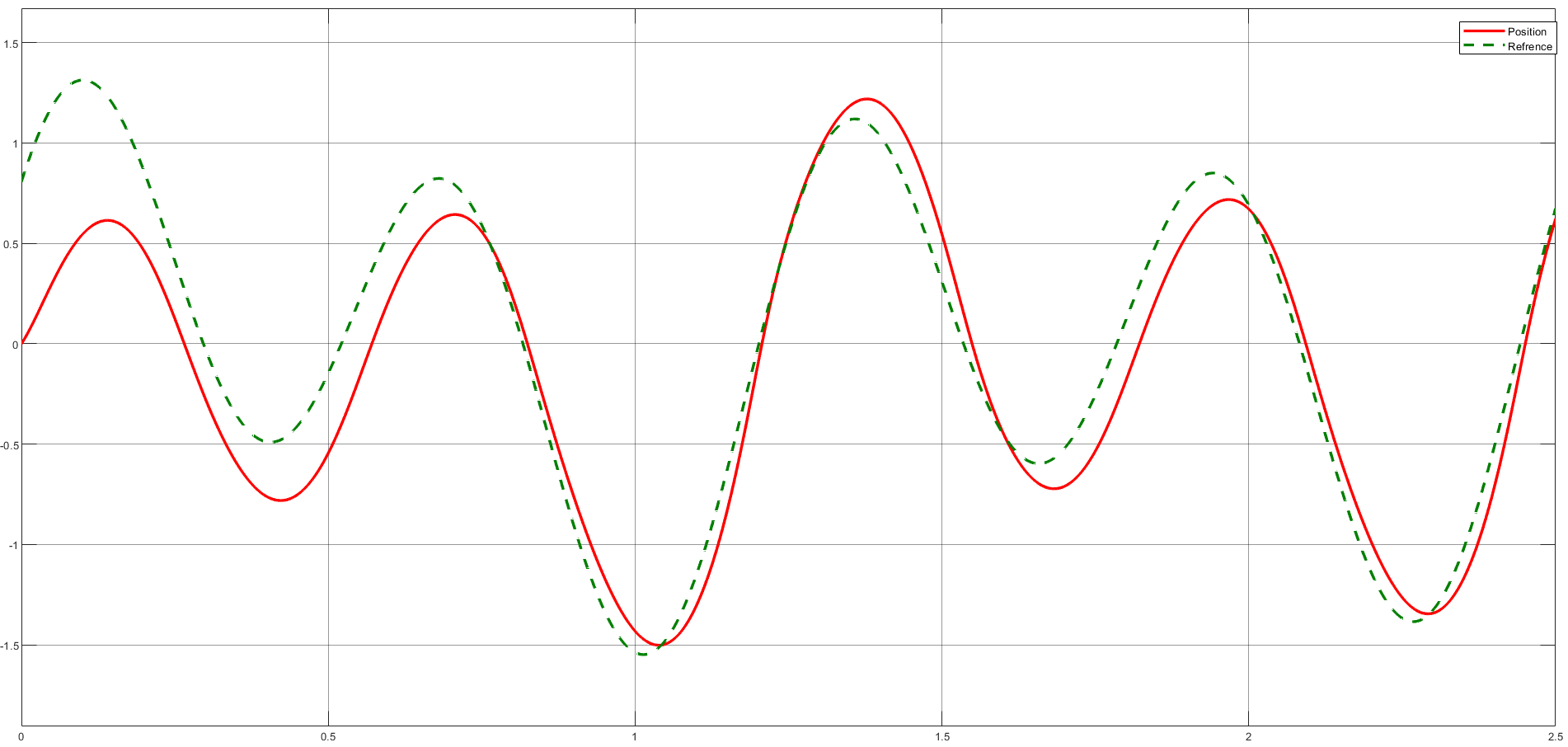}
\caption{Reference signal (dotted green) $5^{th}$ harmonic function with amplitudes of 0.1, 0.5 and 1 and Lyapunov controller output (red) for t=2.5s}
\label{fig:1}       
\end{figure}

\begin{figure}[!t]
\centering
\includegraphics[width=3in]{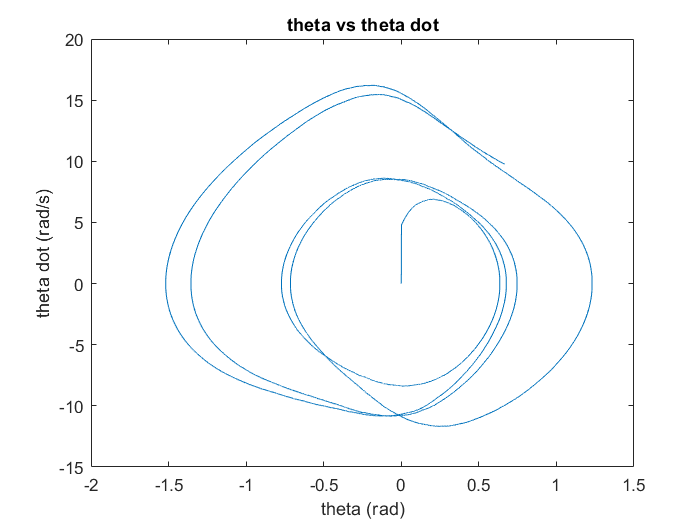}
\caption{Phase portrait showing a stable system with no interference under the harmonic reference signal t=2.5s}
\label{fig:2}       
\end{figure}

A Duffing- van der pol oscillator model with \textdelta=0.5, \textepsilon\textsubscript{1}=1.6, \textepsilon\textsubscript{2}=-0.8, P=3, \textomega=10 and the control parameters \textgamma\textsubscript{1} =12, \textgamma\textsubscript{2} =4, k=115. The parameters were manually adjusted and tuned to give low error \ref{fig:1} and high stability \ref{fig:2}.
\section{LYAPUNOV CONTROLLER PARAMETER OPTIMIZATION}
\label{Section V}
\subsection{OVERVIEW}
The process of selecting the proper Sigmas to reduce the error of the model can be modelled as an exhaustive search problem. Where several Sigmas are tried and the one that results in a reduction in the system error is selected. Though this process results in the best Sigmas, yet it is time consuming. To overcome this problem, we developed a deep learning model that researches different system and controller parameters. A given sigma is good if it reduces the error of the system. The pipeline is presented in Fig. \ref{fig:6}. Initially, the model is run, and the system error is monitored. If the error surpasses a curtain threshold the neural network is queried for a sigma that would reduce the system error given the current state of the system. The algorithm for the entire process is defined below.
The network proposed in Fig.\ref{fig:5} is initially trained and then used to predict new Sigmas; Sigma 1 ($s_1$), and Sigma 2 ($s_2$) if the system error (sys err) exceeds a given threshold (0.8). The network is then given the current state of the system in terms of time (t), position (p), velocity (v), and the randomly generated ($s_1$) and its corresponding ($s_2$) and it predicts the system error. If the predicted error is less than the current system error, then the system is updated with the newly generated Sigmas that are speculated to reduce the error. Every iteration the newly generated Sigmas are their corresponding system parameters are stored. After each 100 iterations the system is probed for perturbations by computing the average of the system error throughout the previous 100 iterations (avg\_sys\_err) if this error is larger than the previous system average (prev sys avg) the network is retrained on a portion of the old data (Memo) used in previous training as well as the data from the previous 100 iterations (new data). The motive behind using the old data as well as the new data is to avoid catastrophic forgetting as mentioned in \cite{Ref14}. \\
 The idea of using a memory is usually known as self-refreshing memory and was initially introduced and tested in \cite{Ref15}\cite{Ref16}\cite{Ref17}. The idea here is that even when there are perturbations they might not last for so long and consequently we don’t want the network to lose its ability to predict the System error if the system parameters return back to norm after the network was retrained several times.
\begin{algorithm}
\SetAlgoLined
Memory = 10\% of the old training data prev\_sys\_avg = system error average of the training data
 initialization\;
 \While{n $>$ 2000}{
 sys\_err = current\_system\_error\;
\If{sys\_err  $>$ 0.8}
	           { \While {True}
	                       {
				$s_1$ = randomly generated\;
	                        $s_2$ = - $s_1$/8\;
	                        predicted\_sys\_err = predict (t, V, p, v, $s_1$, $s_2$)\;
				\If {predicted\_sys\_err  $<$  sys\_err}
	                             {
					Break\;
				      		}
					}
				update\_controller ($s_1$, $s_2$)\;
	                	new\_data = Save (t, V, p, v, $s_1$, $s_2$, new\_sys\_err)\;
	                	avg\_sys\_err = avg\_sys\_err + new\_sys\_err\;}
\If {n == 0}{
	              \If { avg\_sys\_err / 100 $>$ prev\_avg}
	                 { retrain\_network(Memo + new\_data)\;
	                   Memo = Memo + 10\% of the new\_data\;
	                   prev\_sys\_avg = avg\_sys\_err\;
	                   Clear new\_data\;
	                   Clear avg\_sys\_err\;}
}
}
 \caption{Adaptive Algorithm Steps for Sigma prediction}
\end{algorithm}
\subsection{DATASET}
\label{sub1}
The Time, Error, velocity, position and control signal were collected every 1 millisecond and added into an array of values. The dataset was split into 60\% training and 40\% test. A continuous update to the dataset is done according to the calculated error slope every 10ms (100 iterations) as explained above.
\subsection{NETWORK ARCHITECTURE}
\label{sub2}
\begin{center}
 \begin{figure}[!t]
\centering
\includegraphics[width=0.7in]{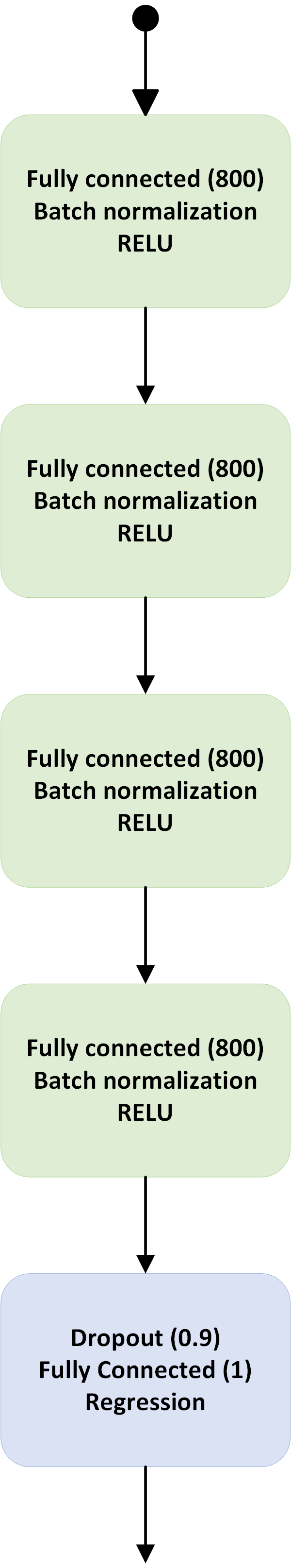}
\caption{Neural network used for predicting the best Sigmas that would result in a decrease in the system error. The input to the network is the time (\textit{t}), displacement (\textit{x}), velocity (\textit{v}), Sigma 1 (\textit{$s_1$}), Sigma 2 (\textit{$s_2$}) and the output is the expected system error (\textit{Error}) for the system. $s_2$ = - $s_1$/8}
\label{fig:3}       
\end{figure}
 \end{center}

The network architecture is in Fig. \ref{fig:3} The network is composed of 5 blocks (15 layers). The number of layers 15 was chosen for its efficiency and high performance. It was found that if a lower number of layers is used the DNN performance was affected and when a higher number of layers was used it had no effect on the performance but reduced the DNN efficiency.  A block is composed of 3 layers: a fully connected convolution layer followed by batch normalization layer and rectified linear unit (RELU) in Eqn. \ref{eq:14}.
\begin{equation}
\label{eq:14} 
output = max (0.0, input)
\end{equation} 
The batch normalization layer allows us to use higher learning rates as it makes ensures that activations are not too high or low. The RELU is a nearly linear function in the sense that it is a piece wise linear function with two linear pieces. This feature increases its capability to preserve many of the properties that make linear models easy to optimize with gradient-based methods as well as the properties that make linear models generalize well. The last block has a dropout layer aside from the fully connected convolution layer and a regression layer to predict the error. Dropout \cite{Ref18} is a regularization method that was developed to solve the problem of overfitting. The overfitting problem occurs when the neural network learns every minor detail in the training data. The effect of such a behaviour is that the network will have high accuracy on the training data but very low accuracy on the test data. This means that the network is unable to generalize on unforeseen data. The proposed solution for this problem is the dropout technique. The dropout technique ignores a number of layer outputs randomly. Such a behaviour makes the layer appear as and be treated-as a layer with a different number of nodes and connectivity to the prior layer. Consequently, each update to a layer during training is performed with a different “perspective” of the configured layer.
 The regression layer computes the mean squared error loss. The network is trained to learn a function F in Eqn. \ref{eq:15}.
\begin{equation} 
\label{eq:15}
F : (t, x, v, s_1, s_2) \rightarrow e
\end{equation} 

Where \textit{s\textsubscript{1}} and \textit{s\textsubscript{2}} are the two Sigmas, \textit{t} is the time and \textit{e} is the error of the model.
\subsection{NETWORK TRAINING AND TESTING}\label{sub3}
\begin{figure}[!t] \centering
\includegraphics[width=3in]{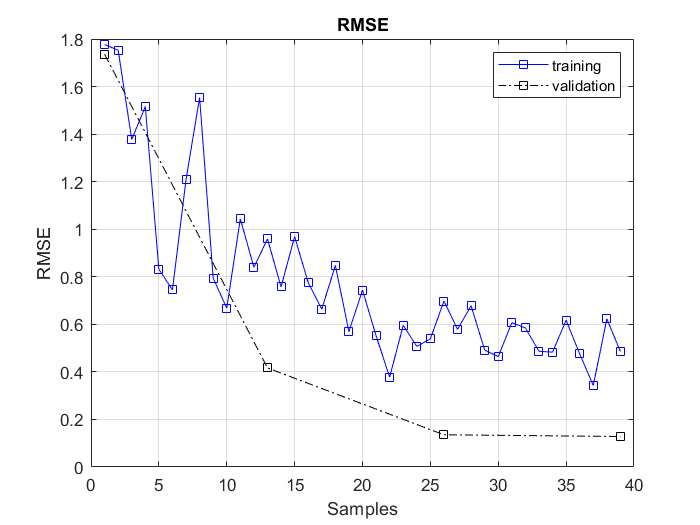}
\caption{Neural network used for predicting the best Sigmas that would result in a decrease in the system error. The input to the network is the time (\textit{t}), displacement (\textit{x}), velocity (\textit{v}), Sigma 1 (\textit{$s_1$}), Sigma2 (\textit{$s_2$}) and the output is the expected system error (\textit{Error}) for the system}
\label{fig:5}       
\end{figure}

The network is initially trained for 5 epochs using Adam optimization algorithm. The learning rate is initially set to 0.001 and is reduced by a factor of 0.2 every 5 epochs. This allows large weight changes in the beginning of the learning process and small changes or fine-tuning towards the end of the learning process. This gives more time for fine-tuning. While the model is running if the system error passes a given threshold the Neural network is used to choose two Sigmas that would lower the error. This is done by selecting a uniformly distributed random number in the interval [-50, 50] for $s_1$. The restriction of $s_1$ and $s_2$ was added later in the research to avoid the genetic algorithm and DNN going into a continuous loop of changing the Sigmas to find the perfect candidate. It was also found that the smaller the sigma the better the system outcome but in order to maintain controller flexibility the range was set to [-50, 50] to account for unexpected changes in the system behaviour. 
The network does not pass the proposed Sigma 1 and Sigma 2 to the model and controller unless Sigma 1 and Sigma 2 are predicted to reduce the system error. This helps in tuning the entire pipeline to take small/large steps to change the behaviour of the duffing van der pol model given that sudden large changes in the Sigmas may result in system failure. 
 The idea behind predicting the error and not the Sigmas is the fact that we can have more than one error for a given Sigma, which gives more flexibility for choosing the range of sigmas to consider.  The root of the mean square error graph for the first 5 epochs are plated in Fig. \ref{fig:5}. The results show that the neural network presented in Fig.  \ref{fig:6} was able to predict the error with an error of 0.03564.

\section{RESULTS AND DISCUSSION}
\label{Section V}
\begin{figure}[!t] \centering
\includegraphics[width=3in]{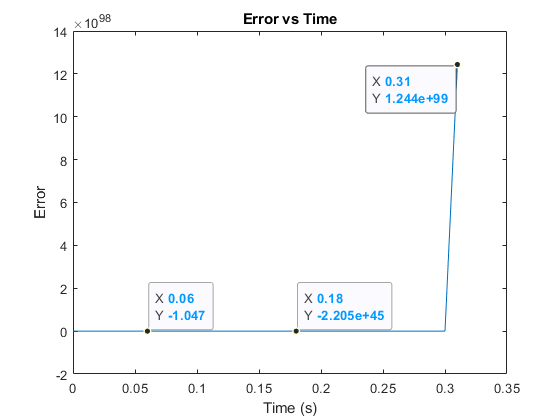}
\caption{The system error goes to infinity as shown when the algorithm is not applied}
\label{fig:6}       
\includegraphics[width=3in]{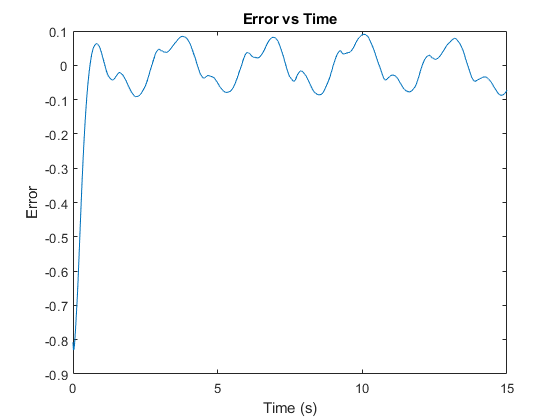}
\caption{The system error after using the neural network in Fig. 7 to get a good estimate of $s_1$ and $s_2$.}
\end{figure}

In Fig. \ref{fig:5} the system is shown to go into instability and fails 0.3 seconds from start due to the use of unfit Sigma 1 = 100 and Sigma 2 = 0.6 while in Fig. \ref{fig:6} we show that using the proposed deep neural network to predict the appropriate Sigma1 and Sigma 2 lead to stabilizing the system and maintaining the system error under the set threshold. The results show the efficiency of the neural network in predicting the error given the sigma, as well as the efficiency of the algorithm in preventing the system from failing likewise in continuously enhancing its performance by keeping the system error as low as possible. On the other hand, maintaining a constant sigma results in large system error and the system might eventually fails.
\subsection{Algorithm Reaction to parameter sabotage}
\label{sub4}
\begin{figure}[!t] \centering
\includegraphics[width=3in]{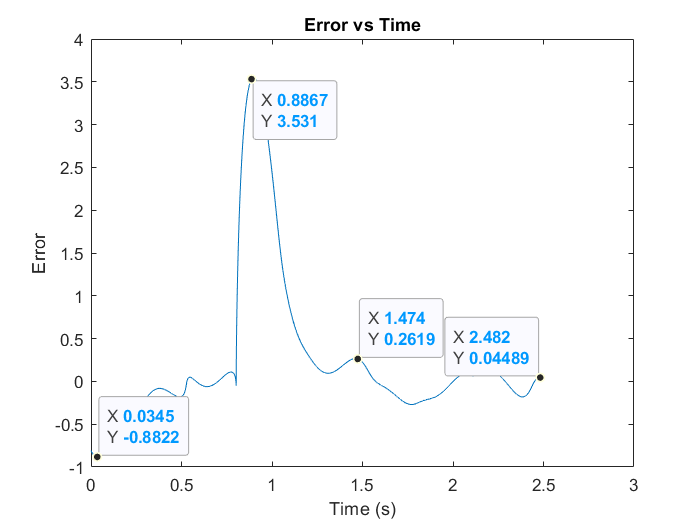}
\caption{Error uptick at t = 0.8867}
\label{fig:7}
\end{figure}

\begin{figure}[!t] \centering
\includegraphics[width=3in]{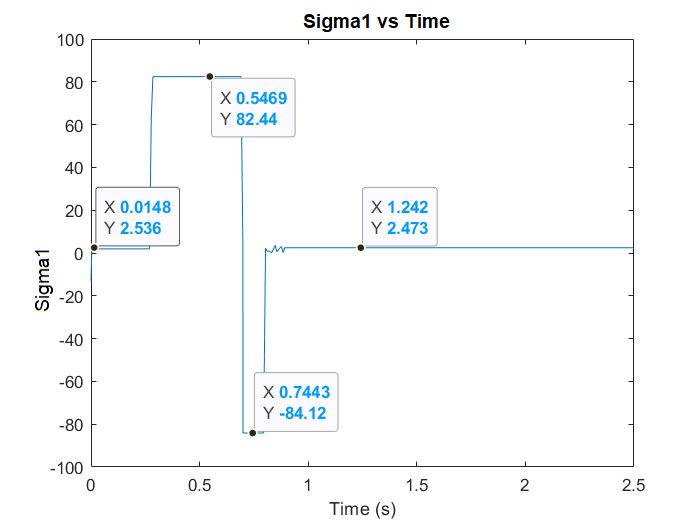}
\caption{The Algorithm reacting to the sudden change by adjusting $s_1$}
\label{fig:8}  
\end{figure}

To elaborate on the ability of the proposed solution to tackle mid system instability or sudden changes. The system parameters are manually overwritten while the system is running to measure the NN and Algorithm effectiveness in returning the system to stability.\\
As demonstrated by Figs \ref{fig:7} and \ref{fig:8} the deep network was able to predict the best parameters to re-establish stability of the system and controller within 0.4 ms. 

\begin{figure}[!t] \centering
\includegraphics[width=3in]{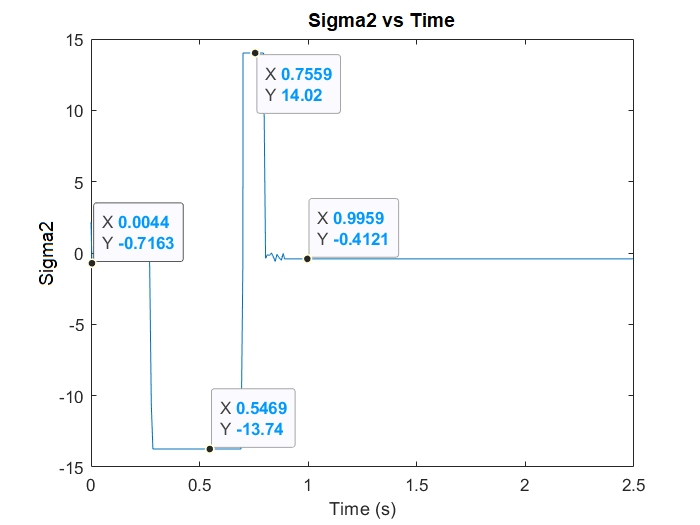}
\caption{The Algorithm reacting to the sudden change by adjusting $s_2$}
\label{fig:9}  
\end{figure}

\subsection{Improving the system performance}
\label{sub5}
In this section we show the ability of the proposed method in finding the best system parameters while the system and controllers are running.
\begin{figure}[!t] \centering
\includegraphics[width=3in]{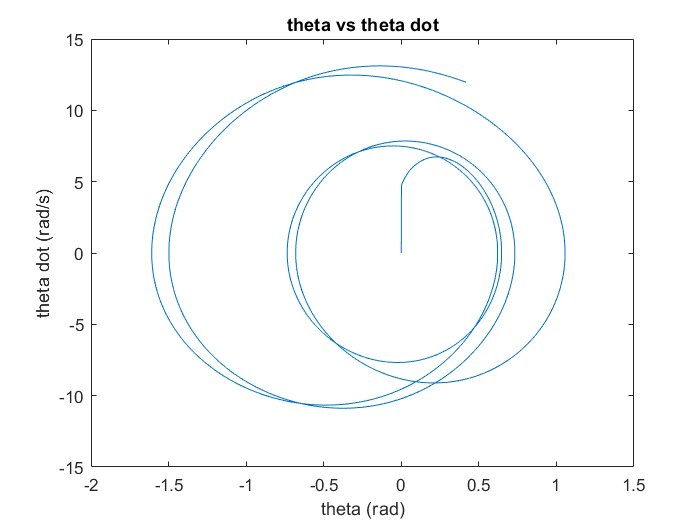}
\caption{Phase diagram of the algorithm before and after finding the optimal system parameters for the conditions which shows major improvement compared to Fig.\ref{fig:5}}
\label{fig:10}  
\end{figure}
\begin{figure}[!t] \centering
\includegraphics[width=3in]{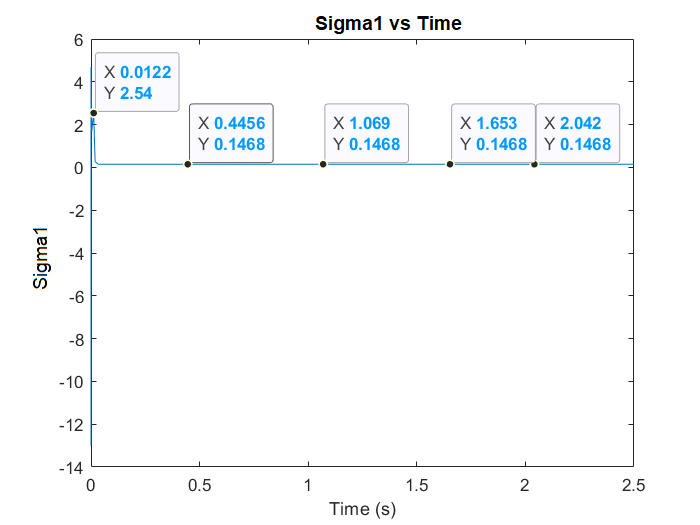}
\caption{$s_1$ and $s_2$ are changed while the system is running to improve performance and reduce error}
\label{fig:11}  
\end{figure}
\begin{figure}[!t] \centering
\includegraphics[width=3in]{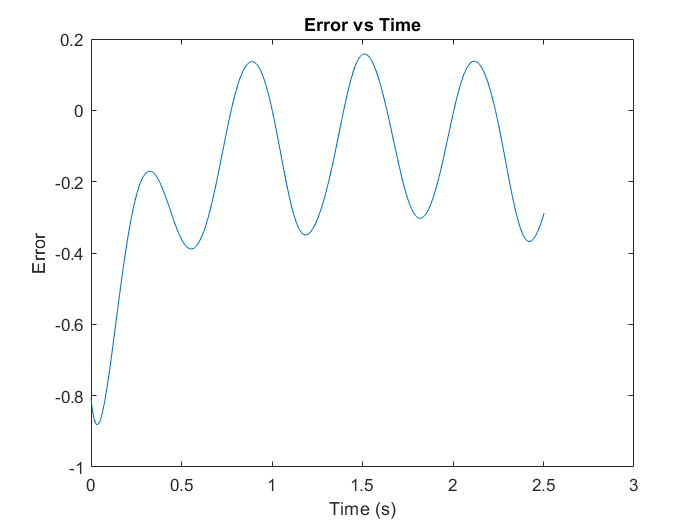}
\caption{Error is shown to be reduced from the starting point by updating the system parameters and the algorithm is successful in maintain the error within bounds}
\label{fig:12}  
\end{figure}
\subsection{Network Retraining}
\label{sub5}
During the re-training phase of the neural network we noticed that without the use of the memory it might take longer and, in some cases, go into an infinite state without finding the optimum solution. The speed with which the system returns back to it stable state affects the retraining phase as we only consider the change in the average of the system error. If the difference between the previous system error average and the new system error average is not large enough then the condition for retraining won’t be met and the system will get stuck trying to find a pair of Sigmas that would result in error reduction. This won’t be possible without memory because the moment the network is retrained the new data overwrites the information learned by the network. The use of memory allows the network to remember the stable and perturbed history of the system parameters.
\begin{figure}[!t] \centering
\includegraphics[width=3in]{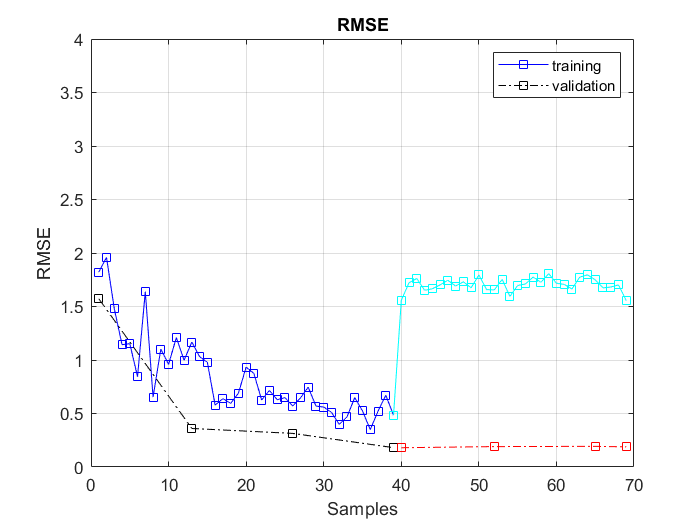}
\caption{Re-training to improve the deep neural network performance in order to adapt to the introduction of new system dynamics}
\label{fig:13}  
\end{figure}
\section{CONCLUSION AND FUTURE WORK}
\label{Section VII}
Lyapunov control was applied to Duffing- van der pol
oscillator model that is experiencing chaotic behaviour. The
study shows the effectiveness of deep learning combination
with nonlinear Lyapunov control in finding the best
parameters to maintain system stability. The study shows
that deep learning with the proposed algorithm enables the
user to effortlessly find the best parameters for the controller
and the system initially and recalibrate the parameters if any
disturbances or new dynamics are introduced to the system.
In future work we would like to investigate the effect of
changing the control strategy according to the type of
instability detected. We speculate that depending on the type
of perturbations and system dynamics, changing the control
strategy might be more effective than changing the
parameters only. Future considerations include a network
that can determine which control strategy would have the
highest impact on returning the system to stability and cause
error reduction.


 \clearpage

\end{document}